\documentclass[11pt,twocolumn]{article}
\usepackage[top=1in, bottom=1in, left=0.78in, right=0.78in]{geometry}
\usepackage{graphicx}
\usepackage{amssymb}
\usepackage{amsmath}
\usepackage{colortbl}
\usepackage{color}
\usepackage[small,bf]{caption}
\usepackage{indentfirst}
\usepackage{url}
\usepackage{subfigure}
\usepackage{multirow}
\usepackage{threeparttable}
\usepackage{tikz}
\usepackage{bbding}
\usepackage[all]{xy}
\usepackage{algorithmic}
\usepackage{array}
\usepackage{multirow}
\usepackage{balance}
\usepackage{paralist}
\usepackage{xspace}
\usepackage{mathptmx}
\usepackage{anyfontsize}
\usepackage{t1enc}
\definecolor{darkgreen}{RGB}{0,0,150}
\usepackage[bookmarks=false, colorlinks=true, plainpages=false,  linkcolor=darkgreen,   citecolor=darkgreen, urlcolor=darkgreen, filecolor=darkgreen]{hyperref}
\usepackage{breakurl}
\usepackage{paralist}
\usepackage[verbose]{wrapfig}

\usepackage[compact]{titlesec}
\titlespacing*{\section}{0pt}{*3}{3.5pt}
\titlespacing{\subsection}{0pt}{*3.5}{3.5pt}
\titlespacing{\subsubsection}{0pt}{*1.5}{0pt}

\usepackage{indentfirst}

\newcommand{\WGS}{whole genome sequencing}

\newcommand{\descr}[1]{\vspace{0.2cm} \noindent \textbf{#1}}

\usepackage[hang,flushmargin,stable]{footmisc}

\date{}
\begin{document}

\title{\bf The Chills and Thrills of Whole Genome Sequencing$^\dagger$}

\author{Erman Ayday$\hspace{0.025cm}^1\;\;\;\;$ Emiliano De Cristofaro$\hspace{0.025cm}^2\;\;\;\;$ Jean-Pierre Hubaux$\hspace{0.025cm}^1\;\;\;\;$ Gene Tsudik$\hspace{0.025cm}^3$\\[1.5ex]
$^1\hspace{0.05cm}$EPFL,\; \{firstname.lastname\}@epfl.ch\\[0.2ex]
$^2\hspace{0.05cm}$University College London\footnotemark[1]~,\; me@emilianodc.com\\[0.2ex]
$^3\hspace{0.05cm}$UC Irvine,\; gts@ics.uci.edu
}

\maketitle

\let\thefootnote\relax\footnotetext{$^\dagger$ A slightly different version of this article appears in IEEE Computer Magazine, Vol. 48, No. 2, February 2015, under the title {\em ``Whole Genome Sequencing: Revolutionary Medicine or Privacy Nightmare.}}

\let\thefootnote\relax\footnotetext{$^*$ Work done while author was with PARC (a Xerox company).}

\begin{abstract}
In recent years, Whole Genome Sequencing (WGS) evolved from a futuristic-sounding research project to an increasingly affordable technology for determining complete genome sequences of complex organisms, including humans. This prompts a wide range of revolutionary applications, as WGS promises to improve modern healthcare and provide a better understanding of the human genome -- in particular, its relation to diseases and response to treatments. However, this progress raises worrisome privacy and ethical issues, since, besides uniquely identifying its owner, the genome contains a treasure trove of highly personal and sensitive information. In this article, after summarizing recent advances in genomics, we discuss some important privacy issues associated with human genomic information and identify a number of particularly relevant research challenges.
\end{abstract}

\section{Introduction}\label{sec:introduction}
Recent years have witnessed impressive advances in  DNA sequencing.
Both throughput and affordability of new-generation sequencing platforms
have increased at a pace faster than Moore's Law would otherwise predict.
It seems quite reasonable to assume that, in a few years, most individuals in developed
countries will have the means of having their genomes sequenced, thus enabling
personalized genomic medicine and facilitating preventive treatment and diagnosis.

However, for now this remains only a prospect and much more research is needed to understand
the very complex relationship between genome and health. To conduct this research, the
scientific community needs large cohorts of patients (or volunteers) willing to share their genetic material.
For instance, the Personal Genome Project involves participants that agree to have their
genomic data -- along  with other personal information -- made publicly available on the Internet,
which raises many potential privacy, ethical, and legal concerns.

The first documented case of privacy issues dates back to the end of the 19th century,
triggered by the availability of a new and revolutionary observation and identification tool:
the photo camera. Since then, several other such tools have become widespread, including:
video cameras, credit cards, Web browsers, and mobile phones. These tools reveal our presence and
habits in various spheres of life, as well as our communication and mobility patterns.
DNA sequencing greatly exacerbates this problem, as the genome represents our ultimate biological
identity. By combining genomic data with information about one's environment or lifestyle
(often easily obtainable from social networks), could make it possible to infer that individual's
phenotype.

In general, access to genomic data prompts some important privacy concerns:
(i) DNA reflects information about genetic conditions and predispositions to specific diseases
such as Alzheimer's, cancer, or schizophrenia, (ii) DNA contains information about
ancestors, siblings, and progeny, (iii) DNA (almost) does not change over
time, hence revoking or replacing it (as with other forms of identification) is impossible, and
(iv) DNA analysis is already being used both in law enforcement and healthcare, thus
prompting numerous ethical issues. Furthermore, it is hard to assess or estimate the extent of the
personal information that could be extracted or derived from the genome in the future.

In this article, after briefly over-viewing some basic genomic concepts, we
describe some expected benefits of personalized medicine and discuss
notable privacy issues, as well as associated research challenges.

\section{Background}\label{sec:genomic_background}
This section provides a brief genomics primer.

\subsection{Processing chain}

The human genome is encoded in double stranded DNA molecules consisting of two complementary
polymer chains. Each chain consists of simple units called nucleotides (A,C,G,T). The DNA of a person can
be retrieved from various sources (e.g., saliva, hair, skin, blood). Once a sample is collected,  the genetic material is extracted and then sequenced -- using a DNA sequencing platform -- to obtain
the so-called raw DNA sequence.
This is usually
in the form of short reads, each including hundreds of nucleotides from random parts of the
genome. Next, the raw reads are quality-controlled, analyzed, and aligned to the reference genome (digital nucleic acid
sequence database, assembled by scientists as a representative example of our species' set of genes), allowing the progressive reconstruction of the whole sequenced genome. The collection of all aligned
raw reads is usually a SAM (sequence alignment/map) file.
There are hundreds of millions of short reads (each including around 100 nucleotides) in the SAM file. Each nucleotide is included in several short reads to have high coverage of each subject's
DNA. After further analysis of the SAM file, eventually, the
approximately 3.2 billion letters in the genome are reconstructed.

The human genome is encoded in double stranded DNA molecules consisting of two complementary
polymer chains. Each chain consists of simple units called nucleotides (A,C,G,T). The DNA of a person can
be retrieved from various sources (e.g., saliva, hair, skin, blood). Once a sample is collected,  the genetic material is extracted and then sequenced -- using a DNA sequencing platform -- to obtain
the so-called raw DNA sequence.
This is usually
in the form of short reads, each including hundreds of nucleotides from random parts of the
genome. Next, the raw reads are quality-controlled, analyzed, and aligned to so-called reference genome (a sequence database, assembled by scientists as a representative example of our species' set of genes), allowing the progressive reconstruction of the whole sequenced genome. After further analysis of these short reads, eventually, the
3.2 billion letters in the DNA sequence of the person are reconstructed.

\begin{figure}[h]%
   \centering
      \includegraphics[height=7cm]{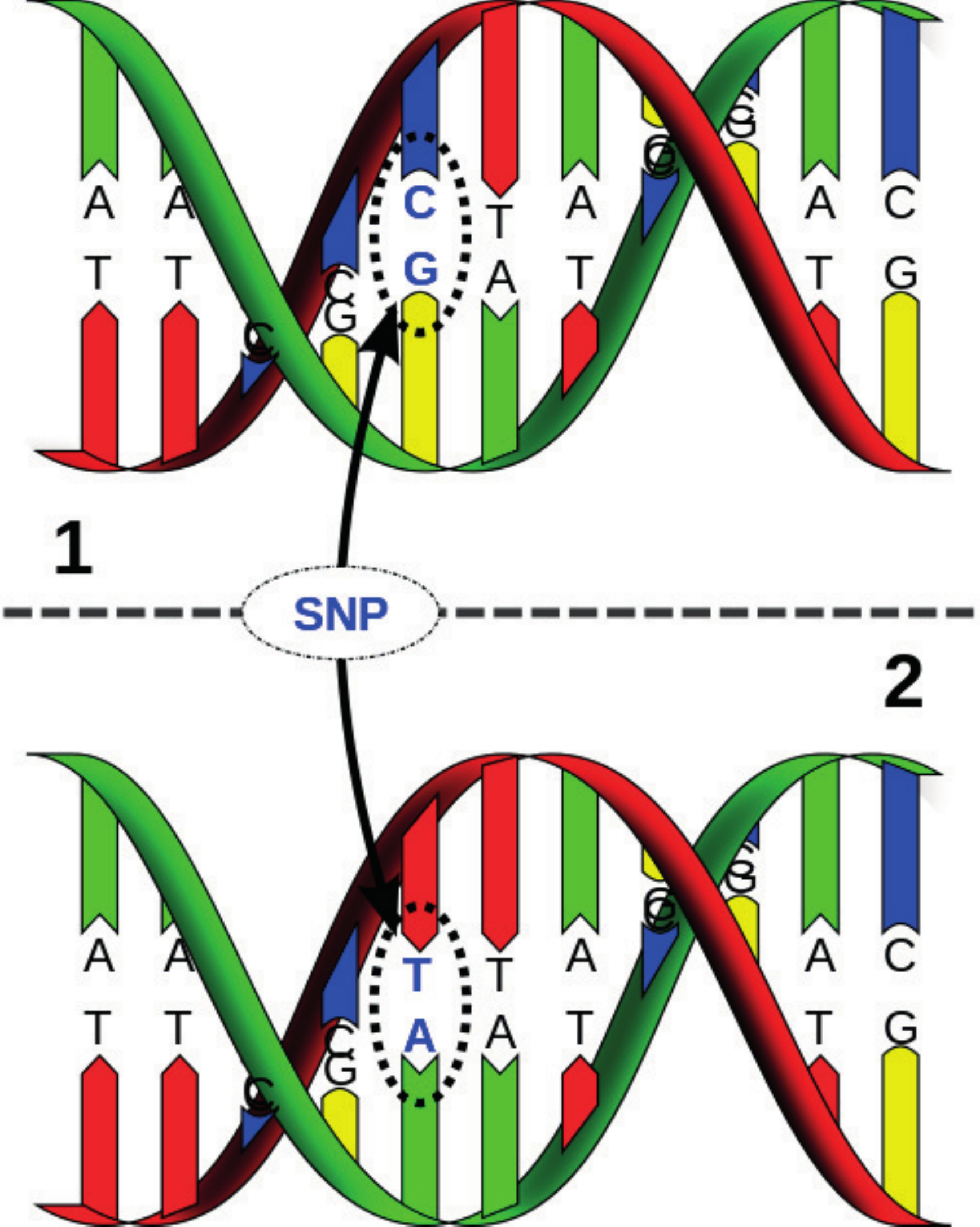}
      \caption{Single Nucleotide Polymorphism (SNP) with alleles C and T.}
      \label{fig:SNP}
\end{figure}

\subsection{Genetic Variations}

Even though most of the DNA sequence is conserved across the whole human population, around $0.5\%$ of each person's DNA
(which corresponds to several millions of nucleotides) is different from the reference genome, owing to genetic variations.
Single nucleotide polymorphism (SNP) is the most common DNA variation. A SNP is a position
in the genome holding a nucleotide that varies between individuals. For example, in Fig.~\ref{fig:SNP},
two sequenced DNA fragments from two individuals contain a single different nucleotide at a particular
SNP position (i.e., locus).
Multiple Genome-Wide Association Studies (GWAS) performed in recent years have shown that a patient's susceptibility to particular diseases can be partially predicted from sets of his SNPs~\cite{Disease_markers_2,Disease_markers_1}. For example, it was reported that there
are three genes bearing a total of ten particular SNPs necessary to (partially) analyze susceptibility
to Alzheimer's disease~\cite{Seshadri_2010}. Thus, leakage of SNPs often poses a significant threat
to individual privacy.
Each SNP contributes to the susceptibility in a different amount and the contribution amount of each SNP is determined by Genome-Wide Association Studies (GWAS)~\cite{GWAS} on case and control groups (these studies are published in several papers).

Each SNP position includes two alleles (i.e., two nucleotides) and everyone inherits one allele of every SNP position from each of his parents. If an individual receives the same allele from both parents, he is said to be \emph{homozygous} for that SNP position. If, however, he inherits a different allele from each parent (one minor and one major), he is called \emph{heterozygous}. It is important to note that a SNP becomes a variant when it carries at least one minor allele.

There are approximately 40 million approved SNPs in the human population as of now (according to the NCBI dbSNP~\cite{NCBI}) and each patient carries on average 4 million variants (i.e., SNPs carrying at least one minor allele) out of this 40 million. We note that the number of approved SNPs in human population is increasing very rapidly~\cite{NCBI}, whereas the number of variants per patient (around 4 million) remains the same. Moreover, this set of 4 million variants is different for each patient.

\begin{figure}[t]%
   \centering
      \includegraphics[width=0.97\columnwidth]{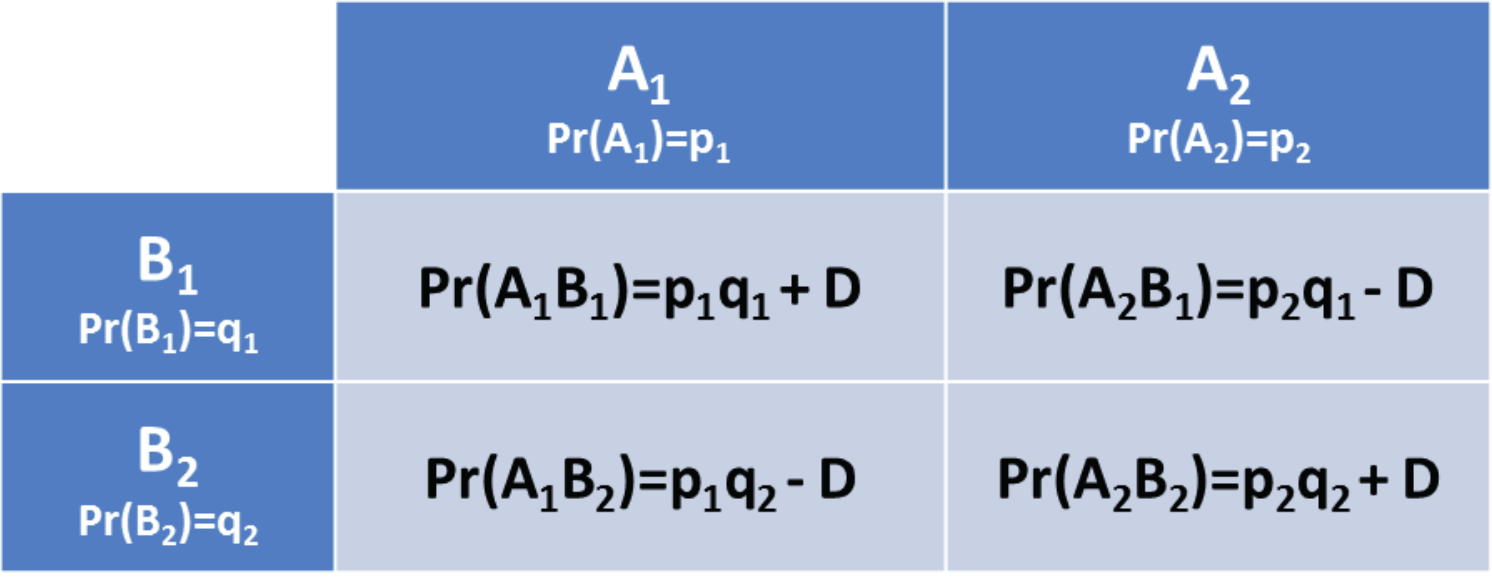}
      \caption{Linkage disequilibrium (LD) between two SNP positions with potential alleles $(A_1,A_2)$ and $(B_1,B_2)$, respectively.}
      \label{fig:LD}
      \vspace{-3pt}
\end{figure}

\subsection{Linkage Disequilibrium}
An interesting characteristic of the SNPs, called \emph{Linkage Disequilibrium} (LD)~\cite{Falconer_book},
poses a notable privacy threat. LD is observed whenever SNPs are not independent
of each other. Therefore, the nucleotide of a certain SNP can be
inferred from the contents of other SNP positions using the LD relationship. The most well-known example of the aforementioned threat is the ApoE status of Jim Watson (the co-discoverer of DNA), who published his genome with the exception of his ApoE gene (which carries SNPs to determine the risk for Alzheimer's disease). However, it was later shown that these SNPs on his ApoE gene can be (probabilistically) inferred using their LD relationships with the published ones~\cite{Nyholt_2009}. For example, assume that $\mathrm{SNP}_i$ and $\mathrm{SNP}_j$ (SNPs which reside at positions $i$ and $j$ on the DNA sequence, respectively) are in LD. Let $(A_1,A_2)$ and $(B_1,B_2)$ be the potential alleles for these two SNP positions (i.e., loci) $i$ and $j$. Further, let $(p_1,p_2)$ and $(q_1,q_2)$ be the allele probabilities of $(A_1,A_2)$ and $(B_1,B_2)$, respectively. That is, the probability that an individual will have $A_1$ as the first allele of $\mathrm{SNP}_i$ is $p_1$, and so on. (Recall that each SNP position includes two alleles, i.e., two nucleotides.) If there were no LD (i.e., if $\mathrm{SNP}_i$ and $\mathrm{SNP}_j$ were independent), the probability that an individual will have both $A_1$ and $B_1$ as the first alleles of $\mathrm{SNP}_i$ and $\mathrm{SNP}_j$ would be $p_1q_1$. However, due to the LD, this probability is equal to $p_1q_1 + D$, where $D$ represents the LD between these two SNP positions. In Fig.~\ref{fig:LD}, we illustrate this LD relationship for all possible combinations of $(A_1,A_2)$ and $(B_1,B_2)$.

\section{Towards Personalized Medicine}\label{sec:good_news}
\begin{figure*}[ht]%
   \centering
   \fbox{\includegraphics[trim = 0mm 5mm 0mm 0mm, clip, height=8.3cm]{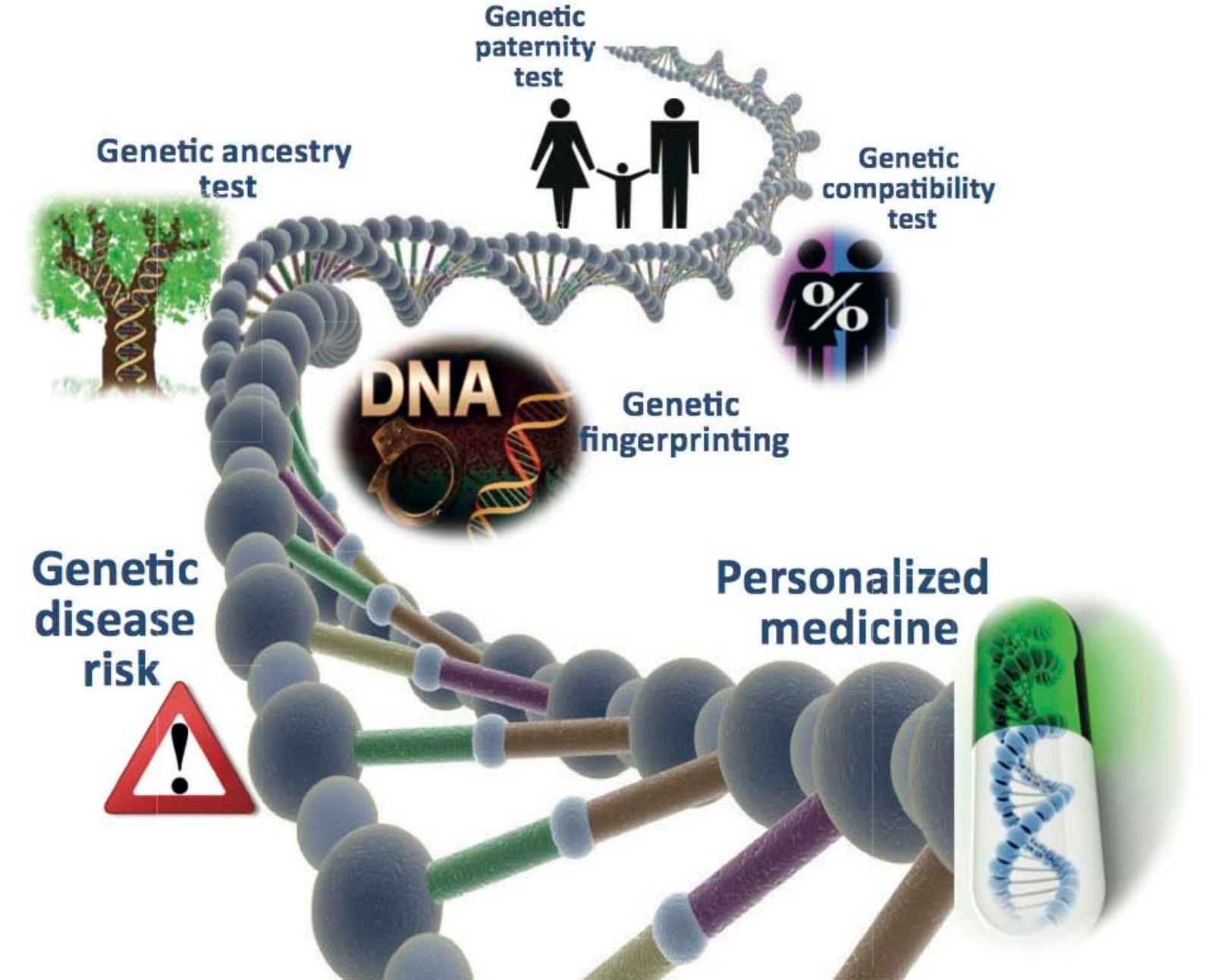}}
      \caption{Applications of genomics.}
      \label{fig:good_news}
\end{figure*}

Widespread and affordable availability of fully sequenced human genomes
creates enormous opportunities, which we summarize in Fig.~\ref{fig:good_news}
(and discuss in this section).

In particular, whole genome sequencing (WGS) facilitates
the advent of a new era of predictive, preventive, participatory,  and personalized medicine (``P4 medicine'')~\cite{hood2009p4}.
\textbf{Personalized genomic medicine} is recognized as a significant paradigm shift and a major trend in health care~\cite{weston2004}, where treatment and medication type/dosage would be tailored to the precise genetic makeup of individual patients.

For instance, certain genetic mutations are known to alter drug metabolism, thus genomic tests are often used today to predict a patient's response to particular drugs. The study of the impact of genetic variations on the response to medications is called {\bf pharmacogenomics}.
A well-known example in this direction includes testing for SNP mutations in the {\em tpmt} gene
for childhood leukemia patients, prior to prescribing 6-Mercaptopurine and Azathioprine drugs.
The {\em tpmt} gene codes for the TPMT enzyme that metabolizes these drugs. Moreover,
genetic polymorphisms affecting enzymatic activity of TPMT are correlated with variations in sensitivity and toxicity response to such drugs. Other common examples include pre-testing for Zelboraf (Roche's treatment for advanced skin cancer), as well as pre-treatment testing for Philadelphia chromosome mutations related to Acute Lymphoblastic Leukemia (ALL) and BRCA1/BRCA2 genes in correlation to familial breast and ovarian cancer syndromes. Experts estimate that about a third of the 900 cancer drugs currently in clinical trials could soon come to market with an enclosed recommendation for a DNA or another molecular test~\cite{900}.

Vanderbilt University's PREDICT program (Pharmacogenomic Resource for Enhanced Decisions in Care and Treatment)~\cite{PREDICT} helps physicians tell which drugs are most likely to work for their patients, and which they should avoid, based on the genetic characteristics of the patients, instead of long trial and error periods. For instance, \cite{PREDICT_article} reports how a specific cholesterol-lowering drug was successfully selected based on the genomic profile of a patient with coronary artery disease.

Experts predict that advances in WGS will further stimulate advances in personalized medicine~\cite{ginsburg2009}. Commercial entities, such as Knome, already offer services that take raw genomic data and create usable reports for doctors. In general, availability of a patient's fully sequenced genome will enable clinicians, doctors, and testing facilities to run a number of complex and correlated genetic tests in a matter of seconds, using specialized computational algorithms, as opposed to more expensive and slower \emph{in vitro} tests.

Another recent Canadian study has shown how, for some cardiac patients, recovery from a common heart procedure can be complicated by a single gene responsible for drug processing, and that selection of blood thinner drugs should depend on whether or not patient holds such a gene mutation~\cite{heart}. Cancer treatment is also one of the most predominant application fields of personalized medicine.

\begin{figure*}[hbt]%
   \centering
   \includegraphics[width=1.4\columnwidth]{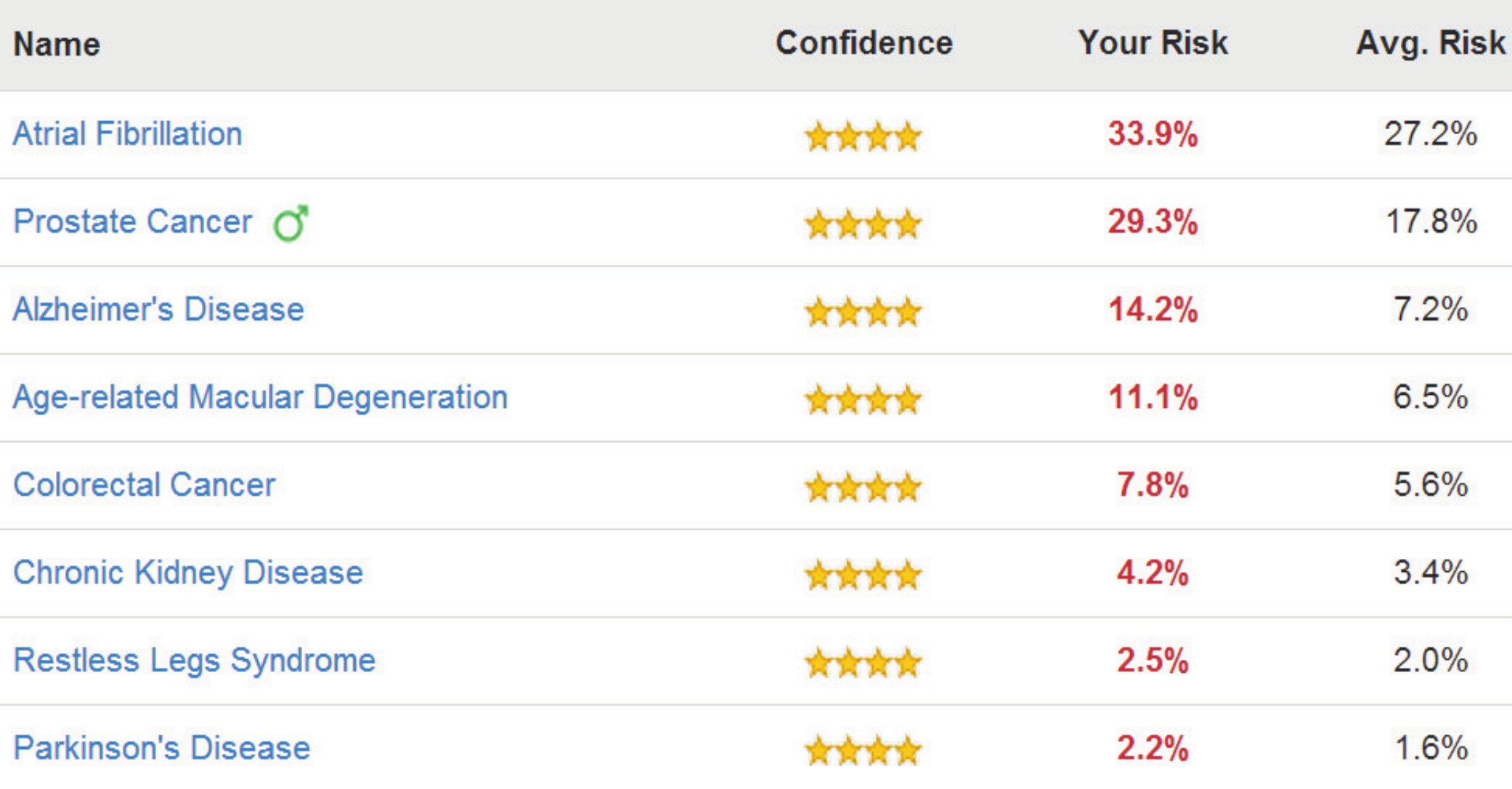}
      \caption{Genetic risks of Greg Mendel for several diseases (source: 23andMe).}
      \label{fig:disease risk}
\end{figure*}

The democratization of low-cost whole genome genotyping and sequencing provides individuals with direct access to their genomic information, including to some \textbf{genetic disease risk tests}. For example, a well-known commercial entity, 23andMe~\cite{23andme}, provides relatively low-cost genetic ancestry and disease risk tests for 960,000 specific SNPs. However, 23andMe does not yet offer WGS. Fig.~\ref{fig:disease risk} illustrates genetic disease risk results of a real human with a fictional name ``Greg Mendel'' provided by 23andMe. It shows the diseases for which Greg Mendel's calculated risk is higher than average. In~\cite{Topol_book}, Topol mentions a few real stories about how the disease risk values obtained from 23andMe helped early diagnosis of serious diseases.

While personalized medicine creates a lot of research ``enthusiasm'', a number of biomedical experts have also
expressed doubts related to the limits of gene mapping's power to predict a person's likelihood of developing a
disease~\cite{fyi}. This is because it is not always clear to which extent certain diseases are correlated to genetic or environmental
factors (a list of diseases known to be associated to genetic features is available in~\cite{Disease_markers_2}).

Availability of WGS will also facilitate faster and lower-cost digital versions of genetic tests that are currently performed \emph{in vitro}. For instance, \textbf{computational  paternity testing} can be designed to mimic its \emph{in vitro} counterpart, with greater speed and accuracy, while preserving its legal acceptance. Furthermore, \textbf{ancestry and genealogical testing} is already offered by several commercial entities. In such tests, publicly available genomic data from individuals belonging to different ethnic groups is compared against the customers' genomic information to understand how the customers relate to known ethnic groups. Similarly, \textbf{genetic compatibility tests}, which let potential or existing partners assess the risk of transmitting to their children genetic diseases with Mendelian inheritance~\cite{mendelian}, are offered by various online services.

\section{Privacy and Ethical Pitfalls}\label{sec:bad_news}
\begin{figure*}[t]%
   \centering
   \includegraphics[width=1.3\columnwidth]{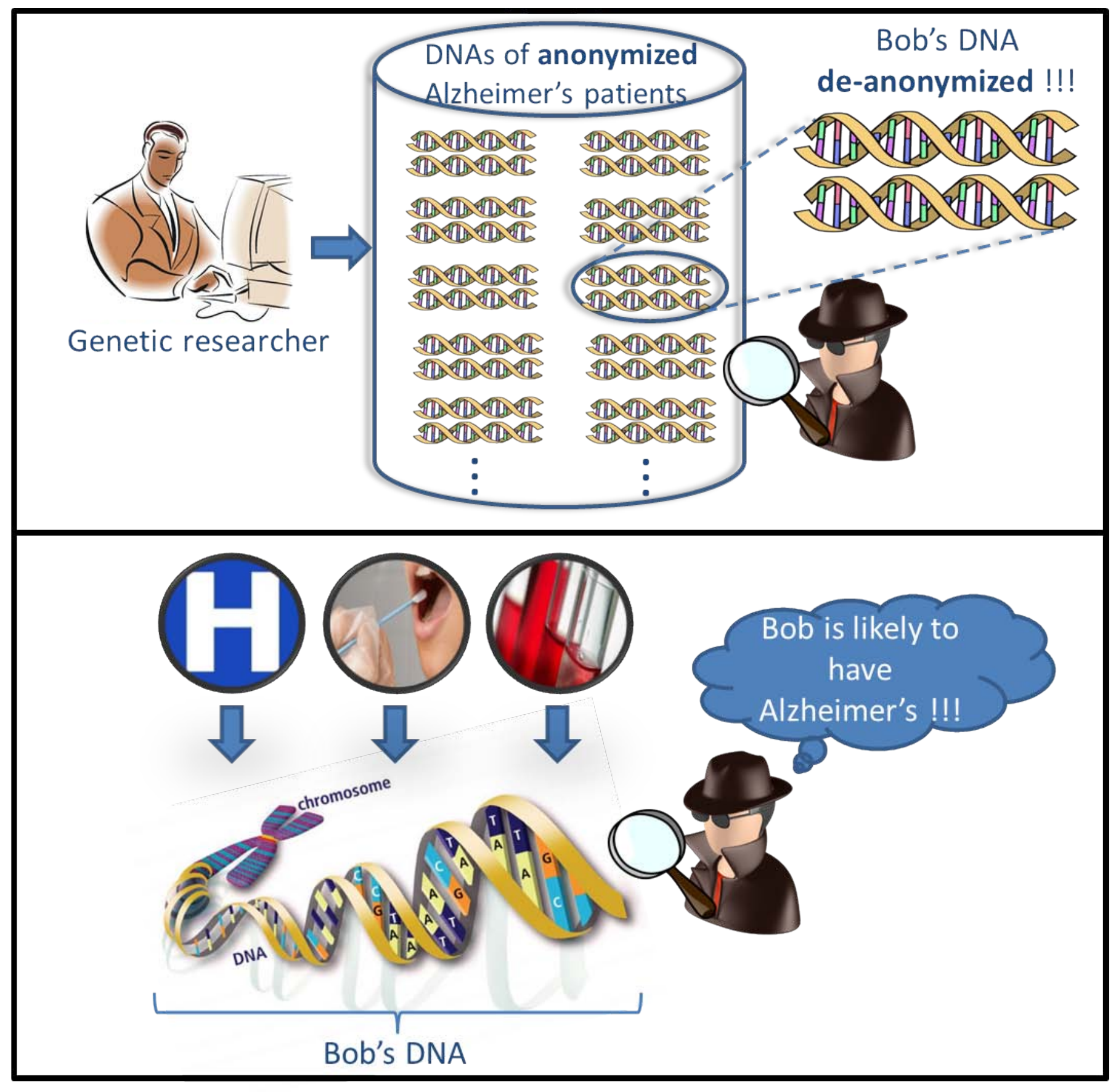}
\caption{Threats to privacy of human genomic data. (Part of the figure is prepared by the U.S. Department of Energy Genomic Science program.)}
      \label{fig:threats}
\end{figure*}

While advances in \WGS\ are paving the way to extraordinary progress
in healthcare (and beyond), they also prompt serious concerns.
Besides  uniquely identifying its owner, a genome contains information about one's
ethnic heritage, predisposition to numerous physical and mental health conditions
as well as other phenotypic traits~\cite{correlated,canli2007emergence,fumagalli2009parasites}. We illustrate two main privacy threats to genomic data in Fig.~\ref{fig:threats}.
Recent studies suggest that even political preferences may be influenced by voters'
genetic makeup~\cite{voters}.

Genomic privacy is often viewed with skepticism, since every individual constantly sheds -- or otherwise
leaves behind -- his biological ``footprints,'' such as hair, skin, or saliva.
This material can be collected (even much later) and used for DNA sequencing.
However, such attacks pose a credible threat only against a targeted individual or a small
group of people. The danger is clearly incomparable with privacy threats posed by access to large numbers
of digitized genomes -- the main focus of this paper.

Furthermore, traditional approaches to privacy, such as de-identification or aggregation~\cite{malin2005},
are ineffective in the genomic context, since the genome itself is the ultimate
identifier~\cite{homer2008}. For instance, a recent study by Gymrek et al.~\cite{gymrek2013identifying} %
demonstrated feasibility of re-identifying DNA donors from a public research database using
information available from popular genealogy Web sites and other available information.
Additional work on genomic re-identification includes~\cite{rodriguez2013complexities} and~\cite{homer2008}.

Moreover, the range of possible abuses is broadened by the increasingly common
handling and sharing of health information (in electronic form) among insurance
companies, health care providers and employers. Unfortunately,
keeping digital records secure is a problem~\cite{med_breach}.
For instance, medical information of $34,000$ patients was leaked from
Howard University Hospital; also, hackers compromised servers of
Utah Department of Health and stolen medical information of almost $800,000$ individuals.

The privacy problem is further exacerbated by the fact that genomes of any two
closely related individuals are highly similar.
Thus, disclosure of one's genome
leads to leakage of significant genomic information about that person's
close relatives, including parents, siblings and offspring. This is a problem
regardless of how the disclosure occurs: voluntarily, accidentally or maliciously.
Therefore, genomic privacy is a unique issue, since, in most other privacy-sensitive
scenarios, only the individual's data is at stake. Whereas, in the genomic
context, disclosure of personal information impacts a potentially large
group of individuals. The most recent example of this issue is the controversy between family members of the deceased Henrietta Lacks (whose genome was sequenced and published after her death, without getting the permission of her family) and scientists who are in favor of publishing genomes online for research purposes~\cite{NYT_lacks}.

Even more worrying is that consequences of genomic data disclosure
are not limited in time. In certain privacy leakage scenarios,
some recourse is possible. For example, bank account numbers and passwords
can be changed, physical or electronic documents (even public key certificates)
can be replaced and old ones can be revoked.
In contrast, a genome is neither mutable nor ``revokable''. Moreover, as large portions thereof are passed on to future generations, disclosure of one's genomic information can turn into an endless curse for future generations.

Based on the above, it is not surprising that privacy concerns represent a formidable
obstacle for assembling large human genomic databases, e.g., for the purpose of conducting
Genome-Wide Association Studies (GWAS).
More generally, privacy concerns might actually stand in the way of advances in medicine and
consequent improvements in overall healthcare. The same could apply in the domain of law enforcement
where DNA-based identification is being increasingly used and there is a need for secure and reliable
handling of large numbers of genomes.

\smallskip
In US, the federal government has been aware of privacy and ethical
issues in genomics. For example,
as early as 1990, the National Human Genome Research Institute (NHGRI) established
the Ethical, Legal and Social Implications (ELSI) Research Program with the
goal of exploring repercussions of advances in genetic and genomic research on
individuals, families and communities.

Federal laws, such as the 2003 Health Insurance Portability and
Accountability Act (HIPAA), provide a general framework for protecting and sharing
Protected Health Information (PHI). Furthermore, the Genetic Information Nondiscrimination Act (GINA)
adopted in 1998 prohibits discrimination on the basis of genetic information
with respect to health insurance and employment \cite{wadman08}. Also, some states, e.g.,
California, have recently started to consider DNA privacy laws~\cite{fear}.

Even the popular culture, via sci-fi movies and literature, has touched upon genetic
discrimination. For instance, the notion of \textbf{genism} that originated in the 1997 movie
``GATTACA'', refers to the theory that distinctive human characteristics and abilities
are determined by genes, resulting in discrimination as pernicious as racism~\cite{annas2001genism}.
Influenced by this movie, a prominent
molecular biologist wrote: ``Gattaca is a film that all geneticists should see if for no other reason
than to understand the perception of our trade held by so many of the public-at-large''~\cite{silver1997genetics}.

While providing general guidelines, current legislation does not offer sufficient
technical information about safe and secure ways of storing and processing
digitized genomes. We believe that security and privacy issues for genomic data
(in the context of both individual genomes and databases thereof) are timely,
important and relatively poorly understood.

Privacy practitioners and consumer organizations are strongly advocating
the \textbf{need for more restrictive legislation} as a result of gaps in current policies.
Also, the Electronic Frontier Foundation (EFF) has analyzed warrant-less
DNA gathering from suspects and against DHS's efforts to collect genetic data from people
placed into administrative detention.
A recent report from the US Presidential Commission for the Study of Bioethical Issues~\cite{president}
analyzed advances of WGS, and highlighted growing  privacy and security concerns.
This report makes a few privacy and security recommendations, including, unfortunately, de-identification.

At the policy level, challenges include the need for informed consent
to guard against surreptitious DNA testing.  Authorities and companies should obtain written permission from citizens
before collecting, analyzing, storing or sharing their genetic information, e.g., preventing
collection of hair or saliva samples and using them for unauthorized sequencing.

On the other hand, some academics fear  that restrictive (privacy-friendly) measures
could seriously hinder genomic research. Scientists typically sequence DNA from large numbers
of people in order to determine genes associated with particular diseases. The informed
consent restriction would mean that large genomic datasets could not be re-used to study a
different disease; researchers would either need to destroy the data after each study, or track down all previously enrolled study participants for each new authorization. %
Also, similarity of related individuals' genomes raises doubts as to whether relatives should also provide consent.

Finally, collection and analysis of human genomes does not arise only in
the contexts of research studies and improved healthcare. It also comes up in increasingly
popular commercial (for-profit) applications, which are not well-regulated.
An example is \url{genepartner.com}, which claims to do matchmaking based on unclear genetic features.

\section{Existing Work on Genomic Privacy}\label{sec:existing_work}
Due to the sensitivity of genomic data, research on the privacy of genomic data has accelerated over the past few years. First, a few techniques have been proposed for secure and privacy-preserving computation on
DNA fragments/snippets. More recently, the security community started focusing on fully-sequenced genomes, motivated by the advances in WGS.

In~\cite{Pastoriza_CCS_2007}, Troncoso-Pastoriza~\emph{et al.} propose a protocol for string searching (then re-visited by Blanton and Aliasgari~\cite{Blanton_dbsec_2010}), where one party with his own DNA snippet can verify the existence of a short template within his snippet by using a Finite State Machine (FSM) in an oblivious manner. Also, secure pattern matching techniques, e.g., those in~\cite{genn-pkc10}
and~\cite{hazayAC10}, have been applied to securely search binary strings in a DNA snippet.
Katz {\em et al.}~\cite{Katz-ccs10} realize
secure computation of the CODIS test~\cite{CODIS}
(run by the FBI for DNA identity testing) and other search tests that could not be otherwise
implemented using pattern matching or FSM.

To compute the similarity of DNA sequences, in~\cite{Jha_SP_2008}, Jha~\emph{et al.} propose techniques for privately computing the edit distance of two strings by using garbled circuits. In~\cite{Bruekers08privacy-preservingmatching}, Bruekers~\emph{et al.} propose privacy-enhanced comparison of DNA profiles for identity and paternity tests, using homomorphic encryption on DNA snippets.

Baldi {\em et al.}~\cite{Baldi_CCS_2011} are the first to focus on whole genomes and
introduce several cryptographic protocols, based on
Private Set Operations, that realize secure testing of whole human genomes,
e.g., paternity tests and genetic screening for personalized medicine or recessive
genetic diseases. In their setting, individuals obtain their genomes
and allow authorized parties (e.g., doctors and clinicians) to run genetic tests
such that only test results are disclosed to one or both parties (with provable security).
In a follow-up work, De Cristofaro~\emph{et al.} proposes a framework and implement a toolkit, called \emph{GenoDroid}~\cite{Emiliano_WPES} for privacy-preserving genomic tests on Android smartphones.
In~\cite{Canim_2012}, Canim~\emph{et al.} propose securing biomedical data using cryptographic hardware.

Ayday~\emph{et al.}~\cite{Ayday_Med_Tech_Report_2012,ayday_NDSS13,ayday_Healthtech13} also focus on the privacy of personal use of genomic data (e.g., in medical tests and personalized medicine methods), and propose methods for protecting user's genomic privacy by considering the statistical relationship between the variants.

When releasing databases consisting of aggregate genomic data (e.g., for research purposes),
known privacy-preserving approaches, such as de-identification, are ineffective on (un-encrypted) genomic data~\cite{Wang_CCS_2009, Malin_boi_2004}. Homer~\emph{et al.}~\cite{Homer_2008} prove that the presence of an individual in a case group can be determined by using aggregate allele frequencies and his DNA profile. In another study~\cite{Gitschier_2009}, Gitschier shows that a combination of information, from genealogical registries and a haplotype analysis of the Y chromosome collected for The HapMap Project, allows for the prediction of the surnames of a number of individuals held in the HapMap database. Thus, releasing (aggregate) genomic data is currently banned by many institutions due to this privacy risk. In~\cite{Zhou_ESORICS_2011}, Zhou~\emph{et al.} study the privacy risks of releasing the aggregate genomic data. They propose a risk-scale system to classify aggregate data and a guide for the release of such data. Recently, Fienberg~\emph{et al.}~\cite{Fienberg_diff_privacy} use differential privacy  to ensure that two aggregated genomic databases, differing from each other by only one individual's data, have indistinguishable statistical features. However, this method does not work well for sparse databases and severely affects the reliability of the genomic data (due to noise injection).

Finally, Wang~\emph{et al.}~\cite{Wang:2009:PGC:1653662.1653703} propose a privacy-protection framework for important classes of genomic computations (e.g., search for homologous genes), in which they partition a genomic computation, distributing sensitive data to the data provider and the public data to the data user.
Also, Chen~\emph{et al.}~\cite{Chen_NDSS_2012} propose a secure cloud-based algorithm to align short DNA sequences to a reference (human) DNA sequence (i.e., read mapping).

\section{Open Research Problems}\label{sec:open_problems}

As discussed above, advances in genomics will soon result in large numbers of individuals
having their genomes sequenced and obtaining digitized versions thereof.
This poses a wide range of technical problems, which we explore below.

\descr{Storage and Accessibility: Genome at Rest.}
Due to its sensitivity and size (about 3.2 billion nucleotides), one key challenge
is where and how a digitized genome should be stored.
It is reasonable to assume that an individual who requests (and likely pays for)
genome sequencing should own the result, as is already the case with any other
personal medical results and information. This raises numerous issues, including:
\begin{compactenum}[{(}1{)}]
\item Should the genome be stored on one's personal devices, e.g., a PC or a smartphone?
 If so, what, if any, special hardware security features (e.g., tamper-resistance) are needed? \smallskip
\item Can it be outsourced to a cloud provider? \smallskip
\item Should the sequencing facility keep an escrowed copy of the genome? \smallskip
\item Should it be entrusted to one's personal physician and/or health insurance provider? \smallskip
\item How is it to be stored: in the clear or encrypted? \smallskip
If the latter, where are encryption keys generated: at the lab? at owner's premises? at the cloud provider? 
 Where are these keys stored? \smallskip
\item How to guarantee integrity and authenticity of the digitized genome? \smallskip
\item Should backups be made? If so, how often and where can copies be kept? \smallskip
\item How can one erase a genome securely? \smallskip
\item Should an individual periodically re-sequence their genome to take advantage 
of more accurate technology?\smallskip
\end{compactenum}

\descr{Privacy: Genome in Action.}
Given the genome's sensitivity, an individual should, ideally, never disclose any
information contained therein. However, this would prevent the access to any genomic application
that cannot be entirely and securely performed {\em in situ}, i.e., within a secure perimeter of one's own personal device.
In principle, this might be possible if operations are performed in some standardized and certified form. For example, if
testing for a genetic disease requires matching a well-known pattern in some
approximate location in the genome, that pattern and its parameters can be certified by
some trusted agency (such as the US Food and Drug Administration). Thus, an individual could be assured that
a legitimate test for a specific genetic disease is being conducted and the result is clearly
communicated to that individual; the latter would then have the option to keep the result private.

At the same time, it is hard to foresee the range and complexity of future genetic
operations: some (future) tests might be too computationally complex to be performed
within the confines of a personal device. Furthermore, some genetic testing would
probably involve multiple genomes, e.g., when tracing origins of some conditions,
siblings or parents/children might need to be tested together. Similarly, in assessing risks of genetic
conditions for future progeny, both prospective parents have to be tested.
Also, some genetic tests constitute intellectual property of a pharmaceutical/biomedical company (which needs to be protected)~\cite{patents,ukpatent,caulfield2006evidence}. %

As soon as genomic information leaves the (virtual) hands of its owner, purely technical
approaches to privacy become insufficient. Legal and professional guidelines are
certainly needed to govern how information is transmitted, stored, processed,
and eventually disposed of on the receiving end, e.g., by the physician, hospital, pharmacist
or medical lab.

\descr{Long-term data protection.} Even if genomes are encrypted, encryption schemes considered strong today might gradually weaken in the long term, whereas genome sensitivity does not dissipate over time.
It is not too far-fetched to imagine that a third-party in possession of an encrypted genome
might be able to decrypt it years or decades later.
For instance, the Advanced Encryption Standard (AES) scheme supports key lengths up to 256 bits --
a key length estimated by NIST, following Moore's law, to be secure several years after 2030~\cite{keylength}. However, computational breakthroughs or unforeseen weaknesses might allow breaking the encryption earlier than expected. Also, even leakage of a long-deceased
individual's genome could affect genomic privacy of that person's living progeny.

Assuming that it can not be copied, an encrypted genome could be periodically re-encrypted. Alternatively, one could split the genome, using secret-sharing techniques,  and partition it among several providers. However, this opens the problem of efficient reassembly of the genome for various operations as well as how to guarantee non-collusion between providers.

\descr{Accuracy and Accountability:}
Computational genomic tests should guarantee accuracy at least equivalent to that of their current
analog {\em in vitro} counterparts. For example, a software implementation of the
paternity test should offer at least the same confidence as
its {\em in vitro} counterpart currently admissible in a court of law.
Also, computational tests should aim at accountability, e.g., by providing lasting
guarantees of correctness for both execution and input information.

\descr{Efficiency.} Computational genomic tests should incur minimal communication/computational costs.
Minimality in this setting is relative to the context of such tests.
For instance, patients may be inclined (and accustomed) to wait several days to obtain
results of genetic tests that concern their health. However in the computational setting,
long running times on personal devices might hinder the real-world practicality of these tests (besides negating one of the main motivations for computational tests.)

\descr{Usability.}
Computational genomic tests that involve end-users should be usable by, and meaningful to, regular
non-tech-savvy individuals. This translates into non-trivial questions, such as:
how much understanding should be expected from a user running a test?
What information (and at what level of granularity) should be presented to the user
as part of a test and as its outcome? Do privacy perceptions and concerns experienced by patients
match those expected by the scientific community? Some users might be willing to forego their genomic privacy in some certain cases. For instance, one may think that patients will be likely to reveal their genomes to their medical doctors (and hence trade off privacy of their genomes) to enable tests that can save them from, e.g., cancer. In contract, in the case of online services or pharmaceuticals, an individual might not wish to forgo privacy. However, very few efforts (e.g.,~\cite{francke2013dealing}) has focused on users' concerns, thus prompting the need for ethnographic studies. Also, there remains an open problem of how to effectively communicate to the users potential privacy risks associated with genomic information and its disclosure.

\descr{Large-scale research on human genomes.} As discussed in Section~\ref{sec:bad_news},
potential privacy, legal, and ethical concerns appear to conflict with
large-scale research on human genomes, such as Genome-Wide Association Studies (GWAS).
However, large scale studies are needed to discover associations  between genetic make-up and medical conditions.
One current trend is to store donors' genomes in the cloud and use analytics
techniques running on powerful computer clusters.  Once again, this prompts
many privacy and legal concerns.

\section{Conclusion}\label{sec:conclusion}
This paper discussed some ``chills and thrills'' of an emerging phenomenon -- affordable and
readily available genomic sequencing. As something radically novel, it brings great opportunities and
significant concerns, especially pertaining to personal privacy. Mitigating
privacy issues will require long-term collaboration among geneticists, other healthcare providers,
ethicists, lawmakers, and computer scientists. As one of the first steps towards such a collaboration, we are involved in organizing a multi-disciplinary seminar on genomic privacy~\cite{dagstuhl}. In order to foster this collaboration, funding agencies need to target this topic. Until recently, at least in the United States, genomic privacy unfortunately fell into a sort of a ``funding gap'' between several agencies. One obvious candidate for playing a key funding role is the National Institute
of Health (NIH). Yet, although it covers both bioinformatics and WGS ethical issues, NIH has funded little research in the
genomic privacy context. The National Science Foundation (NSF), the main agency responsible for
funding academic computer-science research, recently initiated a ``Smart and Connected Health'' program
that includes so-called ``integrative projects'' requiring collaboration among computer
and health sciences. It remains to be seen whether this program will engender long-range
genomic privacy research. Other US funding agencies %
have not, thus far, targeted genomic privacy. A similar situation can be observed in Europe: of course, there are
numerous EU and nationally funded projects focusing on e-health, some of which address
data protection. However, the genomic privacy challenge has been overlooked,
and the number of computer scientists working on the topic is even lower than in the United States.
An additional issue is that, although most officials in charge of data protection typically
have a strong legal background, they lack computer science expertise. Consequently and not surprisingly,
they tend to rely on legislation more than on technology.

In conclusion, we hope that the privacy issues highlighted in the article will be addressed promptly and encourage collaboration among researchers in the fields outlined above. We believe that consideration of such privacy issues will have a positive benefit to society and individuals in their daily lives.

\section{Acknowledgements}
EA and JPH would like to thank Dr. Jacques Fellay, Dr. Paul J. McLaren, Dr. Jacques Rougemont, and Dr. Amalio Telenti for their useful comments and suggestions. They would also like to express their gratitude to Dr. Vincent Mooser, Dr. Didier Trono and Dr. Martin Vetterli for their encouragements in this research endeavor. EDC and GT would like to thank Pierre Baldi, Roberta Baronio, Sky Faber, and Paolo Gasti for their fruitful collaboration.

\balance

{\small
%

\begin{thebibliography}{10}

\bibitem{23andme}
{23andMe}.
\newblock \url{https://www.23andme.com/}.

\bibitem{annas2001genism}
G.~Annas.
\newblock {Genism, Racism, and the prospect of genetic genocide}.
\newblock In {\em World Conference Against Racism, Racial Discrimination,
  Xenophobia and Related Intolerance}, 2001.

\bibitem{Ayday_Med_Tech_Report_2012}
E.~Ayday, M.~Humbert, J.~Fellay, P.~McLaren, J.~Rougemont, J.~Raisaro,
  A.~Telenti, and J.~Hubaux.
\newblock Protecting personal genome privacy: {S}olutions from information
  security.
\newblock Technical report, EPFL-REPORT-182897, 2012.

\bibitem{ayday_NDSS13}
E.~Ayday, J.~Raisaro, and J.~Hubaux.
\newblock Privacy-enhancing technologies for medical tests using genomic data.
\newblock In {\em NDSS (Short Talk)}, 2013.

\bibitem{ayday_Healthtech13}
E.~Ayday, J.~Raisaro, P.~McLaren, J.~Fellay, and J.~Hubaux.
\newblock Privacy-preserving computation of disease risk by using genomic,
  clinical, and environmental data.
\newblock In {\em HealthTech}, 2013.

\bibitem{Baldi_CCS_2011}
P.~Baldi, R.~Baronio, E.~De~Cristofaro, P.~Gasti, and G.~Tsudik.
\newblock Countering {GATTACA}: {E}fficient and secure testing of
  fully-sequenced human genomes.
\newblock In {\em CCS}, 2011.

\bibitem{voters}
D.~Benjamin et~al.
\newblock The genetic architecture of economic and political preferences.
\newblock {\em Proceedings of the National Academy of Sciences}, 109(21), 2012.

\bibitem{Blanton_dbsec_2010}
M.~Blanton and M.~Aliasgari.
\newblock Secure outsourcing of {DNA} searching via finite automata.
\newblock In {\em DBSec}, 2010.

\bibitem{Bruekers08privacy-preservingmatching}
F.~Bruekers, S.~Katzenbeisser, K.~Kursawe, and P.~Tuyls.
\newblock Privacy-preserving matching of {DNA} profiles.
\newblock \url{eprint.iacr.org/2008/203.pdf}, 2008.

\bibitem{900}
A.~Burke.
\newblock {Foundation Medicine: Personalizing Cancer Drugs}.
\newblock \url{http://is.gd/foundation_medicine}, 2012.

\bibitem{Canim_2012}
M.~Canim, M.~Kantarcioglu, and B.~Malin.
\newblock Secure management of biomedical data with cryptographic hardware.
\newblock {\em IEEE Transactions on Information Technology in Biomedicine},
  16(1), 2012.

\bibitem{canli2007emergence}
T.~Canli.
\newblock The emergence of genomic psychology.
\newblock {\em Nature}, 8, 2007.

\bibitem{caulfield2006evidence}
T.~Caulfield, R.~M. Cook-Deegan, F.~S. Kieff, and J.~P. Walsh.
\newblock Evidence and anecdotes: an analysis of human gene patenting
  controversies.
\newblock {\em Nature biotechnology}, 24(9), 2006.

\bibitem{Chen_NDSS_2012}
Y.~Chen, B.~Peng, X.~Wang, and H.~Tang.
\newblock Large-scale privacy-preserving mapping of human genomic sequences on
  hybrid clouds.
\newblock In {\em NDSS}, 2012.

\bibitem{Emiliano_WPES}
E.~{De Cristofaro}, S.~Faber, P.~Gasti, and G.~Tsudik.
\newblock Genodroid: {A}re privacy-preserving genomic tests ready for prime
  time?
\newblock In {\em WPES}, 2012.

\bibitem{Disease_markers_2}
{Eupedia}.
\newblock {Genetically inherited traits, conditions, and diseases}.
\newblock \url{http://www.eupedia.com/genetics/medical_dna_test.shtml}.

\bibitem{Falconer_book}
D.~S. Falconer and T.~F. Mackay.
\newblock {\em Introduction to Quantitative Genetics (4th Edition)}.
\newblock Addison Wesley Longman, Harlow, Essex, UK, 1996.

\bibitem{Fienberg_diff_privacy}
S.~E. Fienberg, A.~Slavkovic, and C.~Uhler.
\newblock Privacy preserving {GWAS} data sharing.
\newblock In {\em ICDMW}, 2011.

\bibitem{correlated}
J.~Fowler, J.~Settle, and N.~Christakis.
\newblock {Correlated genotypes in friendship networks}.
\newblock {\em Proceedings of the National Academy of Sciences}, 108(5), 2011.

\bibitem{francke2013dealing}
U.~Francke, C.~Dijamco, A.~K. Kiefer, N.~Eriksson, B.~Moiseff, J.~Y. Tung, and
  J.~L. Mountain.
\newblock {Dealing with the unexpected: Consumer responses to direct-access
  BRCA mutation testing}.
\newblock {\em PeerJ}, 1, 2013.

\bibitem{fumagalli2009parasites}
M.~Fumagalli et~al.
\newblock Parasites represent a major selective force for interleukin genes and
  shape the genetic predisposition to autoimmune conditions.
\newblock {\em Experimental Medicine}, 206(6), 2009.

\bibitem{genn-pkc10}
R.~Gennaro, C.~Hazay, and J.~Sorensen.
\newblock {Text Search Protocols with Simulation Based Security}.
\newblock In {\em PKC}, 2010.

\bibitem{ginsburg2009}
G.~Ginsburg and H.~Willard.
\newblock {Genomic and personalized medicine: foundations and applications}.
\newblock {\em Translational Research}, 154(6), 2009.

\bibitem{Gitschier_2009}
J.~Gitschier.
\newblock Inferential genotyping of {Y} chromosomes in {L}atter-{D}ay {S}aints
  founders and comparison to {U}tah samples in the {H}ap{M}ap project.
\newblock {\em American Journal of Human Genetics}, 84, 2009.

\bibitem{gymrek2013identifying}
M.~Gymrek, A.~L. McGuire, D.~Golan, E.~Halperin, and Y.~Erlich.
\newblock Identifying personal genomes by surname inference.
\newblock {\em Science}, 339(6117):321--324, 2013.

\bibitem{dagstuhl}
K.~Hamacher, J.~Hubaux, and G.~Tsudik.
\newblock {Dagstuhl Seminar on Genomic Privacy}.
\newblock \url{http://www.dagstuhl.de/13412}, October 2013.

\bibitem{ukpatent}
N.~Hawkins.
\newblock {The Impact of Human Gene Patents on Genetic Testing in the UK}.
\newblock {\em Journal of Genetics in Medicine}, 13(4), 2011.

\bibitem{hazayAC10}
C.~Hazay and T.~Toft.
\newblock {Computationally secure pattern matching in the presence of malicious
  adversaries}.
\newblock In {\em Asiacrypt}, 2010.

\bibitem{homer2008}
N.~Homer et~al.
\newblock {Resolving individuals contributing trace amounts of DNA to highly
  complex mixtures using high-density SNP genotyping microarrays}.
\newblock {\em PLoS Genetics}, 4(8), 2008.

\bibitem{Homer_2008}
N.~Homer, S.~Szelinger, M.~Redman, D.~Duggan, and W.~Tembe.
\newblock Resolving individuals contributing trace amounts of {DNA} to highly
  complex mixtures using high-density {SNP} genotyping microarrays.
\newblock {\em PLoS Genetics}, 4, Aug. 2008.

\bibitem{hood2009p4}
L.~Hood and D.~Galas.
\newblock {P4 Medicine: Personalized, Predictive, Preventive, Participatory A
  Change of View that Changes Everything}.
\newblock \url{http://www.cra.org/ccc/docs/init/P4_Medicine.pdf}, 2009.

\bibitem{Jha_SP_2008}
S.~Jha, L.~Kruger, and V.~Shmatikov.
\newblock {Towards practical privacy for genomic computation}.
\newblock In {\em S\&P}, 2008.

\bibitem{Disease_markers_1}
A.~D. Johnson and C.~J. O'Donnell.
\newblock {An Open Access Database of Genome-wide Association Results}.
\newblock {\em BMC Medical Genetics}, 10(1), 2009.

\bibitem{Katz-ccs10}
J.~Katz and J.~Malka.
\newblock Secure text processing with applications to private dna matching.
\newblock In {\em CCS}, 2010.

\bibitem{malin2005}
B.~Malin.
\newblock {An evaluation of the current state of genomic data privacy
  protection technology and a roadmap for the future}.
\newblock {\em Journal of the American Medical Informatics Association}, 12(1),
  2005.

\bibitem{Malin_boi_2004}
B.~Malin and L.~Sweeney.
\newblock How (not) to protect genomic data privacy in a distributed network:
  {U}sing trail re-identification to evaluate and design anonymity protection
  systems.
\newblock {\em Journal of Biomedical Informatics}, 37, Jun. 2004.

\bibitem{mendelian}
V.~McKusick and S.~Antonarakis.
\newblock {\em {Mendelian inheritance in man: a catalog of human genes and
  genetic disorders}}.
\newblock John Hopkins University Press, 1994.

\bibitem{PREDICT}
{My Drug Genome}.
\newblock {Using Genetics to Personalize Medication Treatment}.
\newblock \url{http://www.mydruggenome.org/overview.php}.

\bibitem{fyi}
G.~Naik.
\newblock {Gene Maps Are No Cure-All}.
\newblock \url{http://on.wsj.com/HackSD}, 2012.

\bibitem{NCBI}
{National Center for Biotechnology Information}.
\newblock {dbSNP}.
\newblock \url{http://www.ncbi.nlm.nih.gov/projects/SNP/}.

\bibitem{patents}
{National Human Genome Research Institute}.
\newblock {Intellectual Property and Genomics}.
\newblock \url{http://www.genome.gov/19016590}.

\bibitem{keylength}
{NIST}.
\newblock {Cryptographic Key Length Recommendation}.
\newblock \url{http://www.keylength.com/en/4/}.

\bibitem{Nyholt_2009}
D.~Nyholt, C.~Yu, and P.~Visscher.
\newblock On {J}im {W}atson's {APOE} status: {G}enetic information is hard to
  hide.
\newblock {\em European Journal of Human Genetics}, 17, 2009.

\bibitem{president}
{Presidential Commission for the Study of Bioethical Issues}.
\newblock {PRIVACY and PROGRESS in Whole Genome Sequencing}.
\newblock
  \url{http://www.bioethics.gov/cms/sites/default/files/PrivacyProgress508.pdf},
  2012.

\bibitem{rodriguez2013complexities}
L.~L. Rodriguez, L.~D. Brooks, J.~H. Greenberg, and E.~D. Green.
\newblock {The Complexities of Genomic Identifiability}.
\newblock {\em Science}, 339(6117):275--276, 2013.

\bibitem{Seshadri_2010}
{S. Seshadri et al.}
\newblock Genome-wide analysis of genetic loci associated with {A}lzheimer
  disease.
\newblock {\em Journal of the American Medical Association}, 303, 2010.

\bibitem{med_breach}
D.~Schultz.
\newblock {Medical data breaches raising alarm}.
\newblock Washington Post, \url{http://preview.tinyurl.com/p9z4q2s}, 2012.

\bibitem{fear}
H.~Shen.
\newblock {California considers DNA privacy law -- Academic researchers fear
  measures would prohibit work with genetic databases}.
\newblock
  \url{http://www.nature.com/news/california-considers-dna-privacy-law-1.10677},
  2012.

\bibitem{silver1997genetics}
L.~Silver.
\newblock {Genetics goes to Hollywood}.
\newblock {\em Nature Genetics}, 17(3), 1997.

\bibitem{NYT_lacks}
R.~Skloot.
\newblock {The Immortal Life of Henrietta Lacks, the Sequel}.
\newblock New York Times, \url{http://nyti.ms/ZiMp2J}.

\bibitem{CODIS}
{The Federal Bureau of Investigation}.
\newblock {Combined DNA Index System (CODIS)}.
\newblock \url{http://www.fbi.gov/about-us/lab/codis}, 2011.

\bibitem{Topol_book}
E.~Topol.
\newblock {\em The Creative Destruction of Medicine: How the Digital Revolution
  Will Create Better Health Care}.
\newblock Basic Books, 2012.

\bibitem{Pastoriza_CCS_2007}
J.~R. Troncoso-Pastoriza, S.~Katzenbeisser, and M.~Celik.
\newblock Privacy preserving error resilient {DNA} searching through oblivious
  automata.
\newblock In {\em CCS}, 2007.

\bibitem{GWAS}
{U.S. Department of Health \& Human Services}.
\newblock {Genome-Wide Association Studies}.
\newblock \url{http://gwas.nih.gov/}.

\bibitem{wadman08}
M.~Wadman.
\newblock Genetics bill cruises through senate.
\newblock {\em Nature}, 453, 2008.

\bibitem{Wang_CCS_2009}
R.~Wang, Y.~F. Li, X.~Wang, H.~Tang, and X.~Zhou.
\newblock {Learning your identity and disease from research papers: information
  leaks in Genome Wide Association Study}.
\newblock In {\em CCS}, 2009.

\bibitem{Wang:2009:PGC:1653662.1653703}
R.~Wang, X.~Wang, Z.~Li, H.~Tang, M.~Reiter, and Z.~Dong.
\newblock Privacy-preserving genomic computation through program
  specialization.
\newblock In {\em CCS}, 2009.

\bibitem{weston2004}
A.~Weston and L.~Hood.
\newblock {Systems biology, proteomics, and the future of health care: toward
  predictive, preventative, and personalized medicine}.
\newblock {\em Journal of Proteome Research}, 3(2), 2004.

\bibitem{PREDICT_article}
K.~Whitney.
\newblock {PREDICT helps pinpoint right statin for patient}.
\newblock \url{http://news.vanderbilt.edu/2012/10/predict-helps-pinpoint}.

\bibitem{heart}
S.~Wood.
\newblock {RAPID GENE: Point-of-care genetic test singles out clopidogrel
  nonresponders}.
\newblock \url{http://www.theheart.org/article/1379471.do}, 2012.

\bibitem{Zhou_ESORICS_2011}
X.~Zhou, B.~Peng, Y.~F. Li, Y.~Chen, H.~Tang, and X.~Wang.
\newblock To release or not to release: {E}valuating information leaks in
  aggregate human-genome data.
\newblock In {\em ESORICS}, 2011.

\end{thebibliography}

}

\end{document}